\documentclass[10pt,twocolumn,letterpaper]{article}

\usepackage{cvpr}              
 
\usepackage[accsupp]{axessibility}

\usepackage{booktabs}
\usepackage{graphicx}
\usepackage{xcolor}
\usepackage{amsmath}
\usepackage{amssymb}
\usepackage{booktabs}
\usepackage{arydshln}

\usepackage[pagebackref,breaklinks,colorlinks]{hyperref}

\newcommand{\fig}[1]{Figure~\ref{#1}}

\newcommand{\tbl}[1]{Table~\ref{#1}}

\newcommand{\xpar}[1]{{\normalfont\bf #1}}

\def\eg{\emph{e.g.}} 
\def\ie{\emph{i.e.}}

\def\etal{\emph{et al.}}

\usepackage{pifont}
\newcommand{\cmark}{\textcolor{teal}{\ding{51}}}%
\newcommand{\xmark}{\textcolor{magenta}{\ding{55}}}%
\newcommand{\cuttmark}{\textcolor{blue}{$\curvearrowleft$}}%

 \newcommand{\supp}{the project webpage\xspace}

 \newcommand{\ours}{VoiceFormer\xspace}

\usepackage[capitalize]{cleveref}
\crefname{section}{Sec.}{Secs.}
\Crefname{section}{Section}{Sections}
\Crefname{table}{Table}{Tables}
\crefname{table}{Tab.}{Tabs.}

\def\cvprPaperID{2204} 
\def\confName{CVPR}
\def\confYear{2022}

\begin{document}
\title{Reading to Listen at the Cocktail Party: \\ Multi-Modal Speech Separation}

\author{Akam Rahimi, Triantafyllos Afouras, Andrew Zisserman\\
VGG, Department of Engineering Science, University of Oxford, UK\\
{\tt\small \{akam, afourast, az\}@robots.ox.ac.uk}\\
{\tt\small \url{https://www.robots.ox.ac.uk/~vgg/research/voiceformer} }
}

\maketitle

\begin{abstract}
The goal of this paper is speech separation and enhancement in multi-speaker and noisy environments using a combination of different modalities. Previous works have shown good performance when conditioning on temporal or static visual evidence such as synchronised lip movements or face identity.
In this paper, we present a unified framework for multi-modal speech separation and enhancement based on synchronous or asynchronous cues.
To that end we make the following contributions:
(i) we design a modern Transformer-based architecture tailored to fuse different modalities to solve the speech separation task in the raw waveform domain;
(ii) we propose conditioning on the textual content of a sentence alone or in combination with visual information;
(iii) we demonstrate the robustness of our model to audio-visual synchronisation offsets; and, 
(iv) we obtain state-of-the-art performance on the well-established benchmark datasets LRS2 and LRS3. 
\end{abstract}

\section{Introduction} \label{sec:intro}

Humans have the remarkable ability to focus on conversations even in a room full of talking people, a phenomenon known as the ``cocktail party effect"~\cite{Moray59cocktailparty}.
Our brains carry out this feat by concentrating their attention to a specific speaker while filtering out the rest of the stimuli originating from interfering voices and other environmental noises.
Although this ability manifests to a limited extent through hearing alone, 
it is greatly enhanced when simultaneous information from other modalities is available.
For example watching a speaker's lips can significantly help disambiguate speech in noise~\cite{schwartz2004b69},
while understanding the natural language context of a sentence enables the listener to anticipate the potential next words of the speaker.

In recent years, progress in audio-visual learning has made it possible for machines to also achieve this ability and very effectively isolate individual voices out of multi-speaker mixtures of speech or noisy audio~\cite{Afouras18,owens2018b,ephrat2018looking}.
Solving this problem enables a great range of practical applications, such as improving subtitle generation in videos with noisy audio, developing smart audio-visual hearing aids to enhance speech conditioned on visual input, or
facilitating teleconferencing in noisy settings such as airports or cars. 

\begin{figure}[t]
  \centering
  \includegraphics[width=0.99\linewidth]{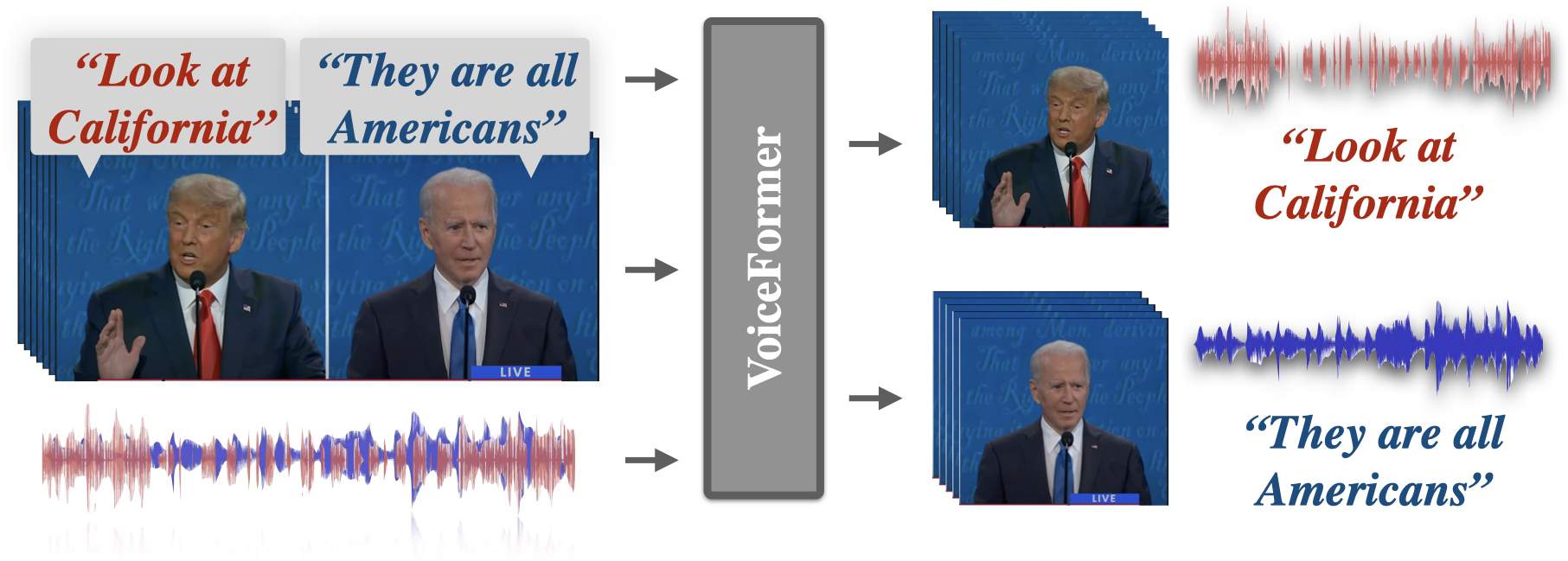}
  \vspace{-10pt}
   \caption{
    We propose \ours, a framework for multi-modal speech separation
    and enhancement, which isolates speech according to
    either the text content of the target speaker's utterance, their lip movements, or both.
    Our framework allows conditioning on cues from multiple modalities, without requiring them to be temporally synchronised or have a common temporal rate. This gives it multiple advantages, such as robustness to temporal misalignments between the inputs.    
   }
   \label{fig:teaser}
   \vspace{-5pt}
\end{figure}

Previous works have principally taken two approaches:  either using synchronous cues, most commonly the lip movements of the target speaker~\cite{Afouras18, ephrat2018looking};
or using static (fixed embedding) cues such as the voice~\cite{wang2018voicefilter} or face characteristics~\cite{chung2020facefilter,gao2021visualvoice}.
The former has the advantage of using dynamic evidence that is very strongly correlated to the desired speech output. However, relying on lip movements has several disadvantages. First, they may be momentarily disrupted -- e.g.\ from visual occlusions -- therefore strong reliance will make the model sensitive to this form of visual noise; and
second, they require synchronisation between the audio and visual streams. 
On the other hand, static cues arising from  the biometric characteristics of the speaker are more robust to temporary disruptions, however, they are not dynamically correlated to the speech (so are a weaker signal) and may be common among different people. For example, it may be increasingly harder to separate speech among individuals with similar voice timbre or appearance.
Recent works have attempted to deal with the inadequacy of conditioning on a single source by either jointly conditioning on more than one modality using a naive fusion of static cues with the lip movement~\cite{Afouras19b} or by learning the separation task jointly with a cross-modal prior~\cite{gao2021visualvoice}.

However, to date, there is no unified framework for: (i) conditioning on {\em asynchronous} information (such as a delay between the audio and visual streams); or (ii) for seamless conditioning (and fusing) on multiple sources of information or on different types of modalities;  or  (iii) for using a large temporal context so that predictions can be made using a language model.

Our first contribution is to enable conditioning on {\em asynchronous} visual (lip) streams. Most previous work relies on costly pre-processing steps to synchronise the audio and video streams, and their performance deteriorates in real-world situations where out-of-sync data is a regular occurrence due to transmission delays, jitter or technical issues.
We show that in our work there are no detrimental effects with timing delays of 5 frames (200 ms)  or more. Furthermore, the audio and visual streams do not even have to have the same temporal sampling rate.

Our second contribution is to enable 
enhancement by conditioning on {\em textual} input. This new functionality allows speech to be enhanced without requiring biometric information or even a visual stream. It is applicable where the textual content of the speech is known in advance, e.g.\ from a prepared speech or lyrics of a song, or where Automatic Speech Recognition (ASR) or lip reading~\cite{Chung16,Assael16} can be used to transcribe what is said, even imprecisely, and then subsequently used to isolate the speaker from background noise. Textual conditioning is asynchronous, as only the order of words (or more precisely the phonemes) is required, but not their precise temporal alignment.

Both of these contributions are facilitated by a new Transformer-based speech separation and enhancement network, where we use the positional encoding of the Transformer to record the timestamp (of the audio and visual samples) or the ordering (of the words in the text) of the conditioning signal.
The network operates directly on the waveform level, without requiring spectrograms as an intermediate step of the audio processing. It uses a U-Net~\cite{ronneberger2015u,defossez2020real} architecture to encode noisy audio and then decode it into clean speech, with the Transformer as the network's bottleneck, where the conditioning information can be visual and/or text. The Transformer also enables modeling
a longer temporal context (\eg~compared to an LSTM) allowing the network to explicitly
model structure in natural language.
By having the ability to anticipate what follows a certain sequence of words the model can then better approximate the target speech output.

In summary, we make the following contributions:
First, we design a modern multi-modal speech enhancement architecture, \ours, that uses a Transformer-based bottleneck to fuse heterogeneous modality streams,
meaning that it can simultaneously condition on multiple non-aligned modalities.
Second, we introduce text-conditioned speech enhancement as a novel multi-modal
task and show that our proposed architecture is well designed to handle it.
Third, we demonstrate that our trained models are robust to audio-visual synchronisation offsets.
Fourth, we exhibit state-of-the-art performance, surpassing other audio-only and audio-visual baselines in the tasks of speaker separation and speech enhancement.



\begin{figure*}[t]
  \centering
   \vspace{-20pt}
   \includegraphics[width=0.9\linewidth]{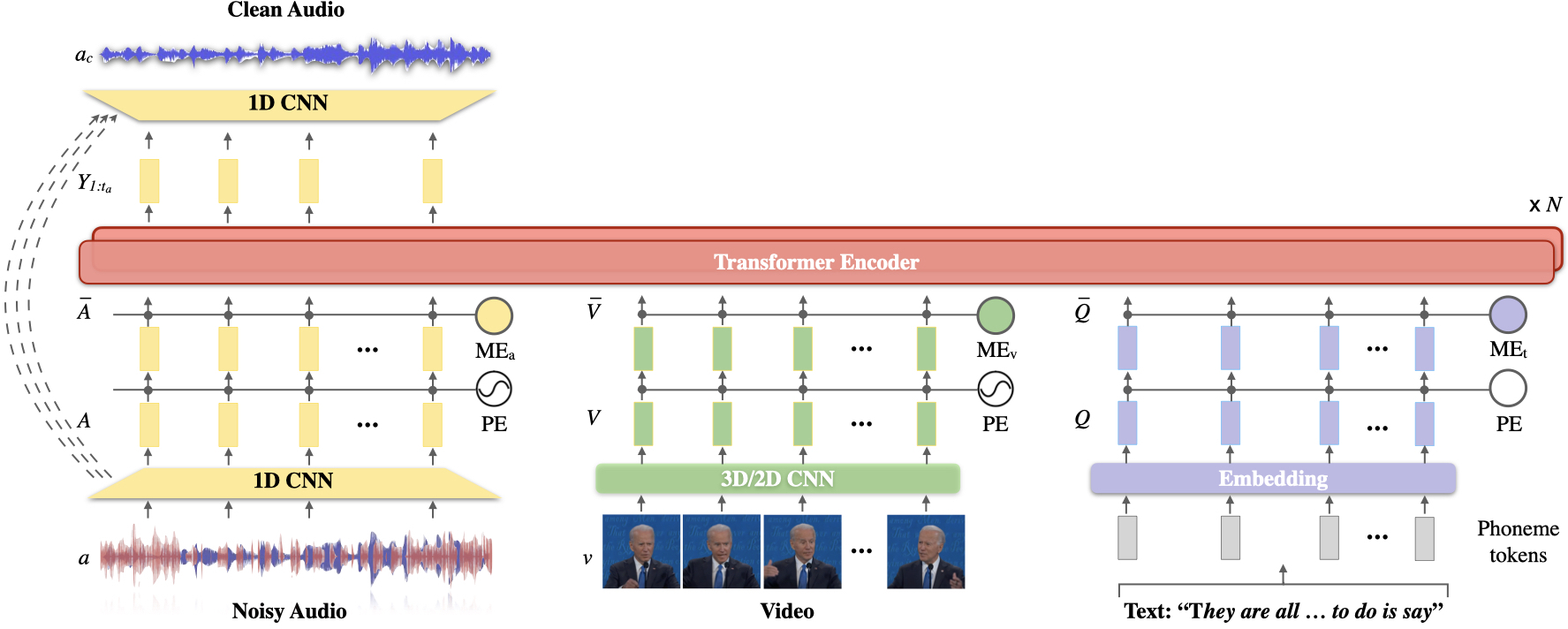}
   \vspace{-5pt}
   \caption{
   Overview of the proposed multi-modal speech enhancement with transformers (VoiceFormer) architecture. it consists of a u-net style encoder-decoder for the audio stream, with the bottleneck layers conditioned on a transformer that can ingest textual and visual modalities. the u-net encoder ingests the raw audio waveform of the target speaker with noise (background or other speakers) and produces a sequence of audio embeddings. the multi-layer transformer conditions on the audio embeddings, the phoneme sequence extracted from the text being spoken, and/or the visual embeddings from the video of the target speaker. the u-net decoder inputs the sequence of refined audio embeddings from the output of the transformer, and produces the clean audio waveform of the target speaker (with the noise removed). In both training and inference, the conditioning can include video or text or both.
   }
   \label{fig:architecture}
   \vspace{-10pt}
\end{figure*}

\section{Related work} \label{sec:related_work}


Our work is related to a large body of previous works which ranges from traditional audio-based speech enhancement to methods for multi-modal speech and sound-source separation.

\xpar{Audio-based speech separation and enhancement.} 
Speech enhancement is a well-studied problem with a long history in audio processing.
Audio-only methods are by design person-agnostic;
they often work well for enhancement, retaining speech and filtering out background noise, but struggle with speaker separation.
Recent methods aim to tackle this by solving the label permutation problem -- assigning audio sources to their corresponding speakers in the audio \cite{Wang,Hershey,Yu}. 
Wang \etal \cite{Wang2019} proposed a method to localize individual speakers
and train an enhancement network  on spatial as well as spectral features.
Lou \etal \cite{Luo2018} introduced a deep learning framework for speech
separation that addresses label permutation and does not require knowledge of the number of speakers.
Yu \etal \cite{Yu} devised a deep learning training criterion
for solving the label permutation problem.
Chen \etal \cite{Chen} perform separation based on clustering
of the audio sources in the embedding space.
Defossez \etal~\cite{defossez2020real} recently proposed the Denoiser, 
a real-time speech enhancement network that is trained end-to-end on raw waveforms.

\xpar{Multi-modal methods based on static cues.} 
Various recent methods attempt to solve the audio separation problem based on
external cues that contain information about the sound source.
Examples of such methods are works that perform identity-specific 
speech separation using voice~\cite{wang2018voicefilter}, or face~\cite{chung2020facefilter} identity embeddings. 
Related works that also fall in this category are various audio-visual
methods for separating the sounds of musical instruments based
on stationary appearance cues~\cite{gao2019co,gao18graum,zhao2018sound,rouditchenko2019self,fisher2000learning,Jie17,Xu19plusnet,tzinis2020into,Tzinis2021ImprovingOS}.

\xpar{Multi-modal methods based on dynamic cues.} Another family of methods solve the source separation task based on dynamic cues that exhibit some variation over time. Such cues are more commonly contained in a synchronised visual stream, therefore these methods are in their majority audio-visual.
For example, utilising visual features has proven to be very beneficial in
separating speakers in audio clips where the corresponding video is accessible.
Indeed, recent works~\cite{Afouras18,ephrat2018looking,Afouras20b,gabbay2018seeing,gabbay17b} conditioned
deep learning frameworks on the lip movements of target speakers in order to
isolate their voice among multiple other speech signals.
Wu \etal \cite{Wu} presented an audio-visual speech separation network that operates in the time-domain (raw waveforms) instead of
frequency-domain, while Sadeghi~\etal
solved the task by employing audio-visual VAEs \cite{sadeghi2020audio}.
 Owens \etal~\cite{owens2018b} use spatio-temporal
video features for separating on-screen from off-screen speech.
In a similar line of work,
recent methods have proposed using motion 
cues in videos~\cite{Parekh17,zhao2019sound,Gan2020MusicGF} in order to separate musical instruments belonging to the same class, thus overcoming the limitations of static appearance features.   
A few recent works investigated combining static and dynamic cues
to improve speech separation.
For example,  Gao \etal~\cite{gao2021visualvoice} investigated 
training an audio-visual speaker separation network jointly with voice-face embeddings that provide a prior to aid the separation process. 
Another proposed direction~\cite{Afouras19b,gu2020multi} is conditioning a separation network on both lip-movements and  
an embedding of the target speaker's
voice to improve robustness to visual occlusions.
However, none of those works propose a unified framework for conditioning on multiple non-aligned dynamic sources of information.

\xpar{Multi-modal fusion.}
Our method is more broadly related to works that fuse different modalities to solve multi-modal tasks, such as audio-visual fusion using Transformers~\cite{jaegle2021perceiver,nagrani2021attention} for audio-visual event detection~\cite{Lin_2020_ACCV,tian2020avvp,lee2021parameter} or audio-visual synchronization\cite{Chen21b}, 
video-text fusion for visual grounding~\cite{xu2019multilevel,yuan2019find} and visual keyword spotting~\cite{Prajwal21}.


\section{Method} \label{sec:method}

In this section, we describe our proposed method for multi-modal speech enhancement which we call the \ours. Given a noisy speech signal, the goal is to separate a target speech component corresponding to other input modalities (text or video), and to filter out the rest of the signal (other speakers or background noise).  
An overview of the architecture is shown in~\fig{fig:architecture}.
We use a U-Net style audio encoder-decoder (similar 
to~\cite{defossez2020real}),
with a multi-modal Transformer in its bottleneck, where the noisy audio is
fused with the conditioning inputs (video and text).
The rest of this section describes the individual components and the training of the model.
We refer the reader to \supp for full architectural details.

\subsection{Architecture}

\xpar{Audio, Visual, and Text representations.}

The model has three input streams: one ingesting an audio waveform $a\in \mathbb{R}^{T_a}$,
one the corresponding video input
$v\in \mathbb{R}^{3\times T_v \times H \times W}$,
and one a textual representation $s = (s_1, s_2 \cdots, s_{n_{s}})$ of the sentence being uttered.

Similar to ~\cite{defossez2019}, we extract a representation of the noisy audio, $A\in\mathbb{R}^{t_a \times c}$, directly from the input waveform $a$ using the encoder part of a U-Net, which consists of 1D convolution layers. 
We also use a VTP network ~\cite{Prajwal21} to obtain visual representations, $V\in\mathbb{R}^{t_v \times c}$.
The textual representation is obtained using the Phonimizer library \cite{phonimizer} with espeak-ng as its backend;
the words in the input sentence are first mapped to a phonetic sequence of length $t_q$ based on the International Phonetic Alphabet and are then mapped to a sequence of learnable embedding vectors $Q \in \mathbb{R}^{t_q \times c}$.

\xpar{Transformer bottleneck.}
In order to inform the model of the temporal order of its inputs, \ie~signal timestamps for the video/audio features and phoneme ordering for the text,
we add positional encodings, 
$PE_{\{a,v,q\}} \in \mathbb{R}^{t_{\{a,v,q\}} \times c}$.
$PE_{a}$ and $PE_{v}$ are implemented as sinusoidal vectors 
and $PE_{q}$ as learnable embedding vectors.
Moreover, in order to allow the model to distinguish which signal comes from which modality, we also add
modality encodings, 
$ME_{\{a,v,q\}} \in \mathbb{R}^{c}$, which are three learnable vectors, one for each modality type. In summary, the order and modality aware uni-modal representations are calculated as 
{
\small
\begin{align}
&\overline{A} = A + PE_{a} + ME_{a} ,  \\
&\overline{V} = V + PE_{v} + ME_{v} , \\ 
&\overline{Q} = Q + PE_{q} + ME_{q} ,  
\end{align}
}%
These are concatenated along the temporal dimension 

{
\small
\vspace{-10pt}
\begin{align}
Z = ( \overline{A}; \overline{V}; \overline{Q} ) \in \mathbb{R}^{ ( t_a + t_v + t_q)\times c}
\label{eq:concat}
\end{align}
}%
and the resulting feature vector is processed with a Transformer encoder with $N$ layers and $h$ heads:
\[
Y = \textsc{Transformer-Encoder}( Z ).
\]
The Transformer bottleneck fuses the three inputs together, allowing for full cross-attention between all modality combinations.
In particular, the textual and video evidence is attended upon and used to extract only the relevant parts of the embedded audio. 

We note that neither explicit alignment nor common frame rates between the different signals are required. 
The output of the Transformer corresponding to its audio input, $Y_{1:t_a}$, contains
the representation of the separated/enhanced audio (the outputs corresponding to video and text are discarded).

\xpar{U-Net Decoder.}
The enhanced audio representation is decoded into a waveform $\hat{a}_{c}$ with the decoder part of the U-Net, which is comprised of a stack of transposed 1D convolutions, including shortcut connections from the audio encoder. 
The resulting output $\hat{a}_{c}$ is an enhanced waveform containing only the speech corresponding to the visual and text input.

\subsection{Training objective}
Given a training dataset $\mathcal{D}$ of tuples ($a,v,s,a_c$) of noisy audio waveforms and corresponding video, text and clean target waveforms, we train the model using an L1 loss between the predicted enhanced and the target clean waveforms:
\begin{align}
\vspace{-10pt}
\mathcal{L} = 
\mathbb{E}_{(a, v, s, a_{c}) \in \mathcal{D}} 
\ \| a_{c} - \hat{a}_{c} \|_1
\end{align}

\section{Experiments} \label{sec:experiments}

\subsection{Synthetic sequences}
Following previous work~\cite{Afouras18,ephrat2018looking} we train and evaluate our models by creating
synthetic noisy samples by adding the waveforms of two separate clips and requiring the model to
reconstruct the individual waveforms in the output, after conditioning on the corresponding
video/text input. 
In particular, one of the clips always contains clean speech of a single speaker, while the interfering audio might be either speech from another speaker in the speaker separation experiments, or a noise audio clip, simulating background noise for the speech enhancement experiments.
Note that although we train and evaluate the model on synthetic audio mixtures it is applicable to real noisy sequences as the domain gap between synthetic and real samples is small. 

\subsection{Datasets, training \& evaluation protocol}
 \xpar{Data.} 
We obtain audio-visual speech samples from the LRS2~\cite{chung17} and LRS3~\cite{Afouras18d} lip reading datasets.
LRS2 contains broadcast footage from British television while LRS3 has been created from TED and TEDx clips downloaded from YouTube. 
Both datasets contain audio-visual tracks of tightly cropped talking heads, engaging in continuous speech. All tracks are accompanied by text transcriptions of the utterances that are well aligned to the video and audio, which have also been automatically synchronised.
On examining the datasets we determined that some of the samples include two speakers while others included background noise such as clapping, bird chiming, music or crowd laughter. These samples were removed from the dataset such that every sample contains speech from only one speaker without other background noise. A combination of diarization and background noise detection methods were employed to detect and remove the noisy samples. As a result, 57 hours out of 197 hours was retained from the LRS2, and 439 hours out of 440 hours was kept from LRS3.

Moreover, for obtaining noise for our denoising experiments we  follow~\cite{defossez2020real} and use a subset of the DNS \cite{reddy2021interspeech} dataset, which contains approximately 181 hours of noise audio from a wide variety of events.
These samples were used as background noise to construct synthetic noisy audio waveforms during training and evaluation. 

\xpar{Training sequences.}
Approximately 23,000 samples from the LRS2 dataset and over 100,000 samples from the LRS3 dataset were used at training time. The samples were mixed together randomly at each training pass  leading to the generation of numerous new and unseen examples. 
The speech signals are mixed together in sequences of 4 seconds for the version of the model trained on audio and video input. The starting point of each sequence was randomly chosen as a data augmentation method. The length of the sequences for the models that have text as their input is dictated by the number of words in the text that correspond to the associated audio and videos, ranging between 1 to 6 seconds. The samples with similar lengths were batched together at training time to avoid padding the sequences with zeros whenever possible. Each audio track was independently normalised before mixing them together to create synthetic mixtures or feeding them into the network. 


\xpar{Evaluation sequences.}
We evaluate on synthetic mixtures of two speakers, or a mixture of one speaker and a noise sample. To distinguish between these two related tasks we refer to the first one as speaker separation and the second as denoising.

We created separate test sets from LRS2 and LRS3 with $2515$ and $3229$ samples from each dataset respectively. These test sets were used to evaluate the aforementioned tasks, comparing our model with baselines and for performing model ablations and robustness tests.
The samples in all the test sets contain noisy audio, video and text. The duration of the samples varies as they are cropped based on the length of the corresponding text. The text samples are 9 words long and do not necessarily form a sentence. 
We include qualitative examples of real sequences on \supp.

\xpar{Evaluation metrics.} For evaluating our methods and baselines we use standard speech enhancement metrics,  
including Signal-to-Distortion-Ratio (SDR) \cite{fevotte05,Raffel14mireval}, a common blind source separation criterion, measuring the
ratio between the energy of the target signal and of the errors
contained in the separated output, Short-Time Objective Intelligibility (STOI)~\cite{taal11}, which measures the intelligibility of the signal, and the Perceptual Evaluation of Speech Quality (PESQ)~\cite{rix2001perceptual}, which rates the overall perceived quality of the output signal.   


\subsection{Implementation details}

Our network is implemented and trained in Pytorch. The faces in the video recordings are cropped and resembled into 25 FPS. The audio input is converted into mono by taking the mean of both channels and resampled to have a  rate of 16kHz, and the signals are upsampled by 3.2 times before feeding into the network. The audio output of the model is down-sampled by the same ratio. 
For the audio U-Net, we use the Denoiser implementation of ~\cite{defossez2020real} without any changes to the architecture.  
For the visual backbone, we follow ~\cite{Prajwal21} and use a VTP network consisting of transformer layers on top of a 3D/2D residual CNN~\cite{Stafylakis17}, pretrained on a word-level lip reading task. To speed up training, the backbone is frozen and visual features are pre-extracted and saved on a hard drive.
For the Transformer encoder, we use $N=3$ layers and $h=8$ heads, with a model size of $532$. The embedding dimensions across all the modalities is set to 768 to match the channel dimension of the audio features from the output of the U-Net encoder layer. 

We obtain models that condition on a single modality (e.g.\ only video or only text), by simply not including the corresponding input in Eq.~\ref{eq:concat}.

Training started with a network that includes an LSTM on synchronised audio and corresponding visual sequences as a pre-training stage for the Transformer model. For training the transformer the learning rate was set to $5 \cdot 10^{-5}$. In all cases the Adam optimizer was used with a weight decay of 0.0001, batch size of 64 and the learning rate was reduced linearly after each epoch on plateau. The models are trained on the LRS2 and LRS3 datasets separately starting with the LRS2 mixtures.
The training curriculum for the speech enhancement models started with mixtures of 2 speakers before training on one speaker and background noise. 
When experimenting with more than two transformer layers, all the trainable parameters in the model were frozen apart from the additional encoder layers, which helped to stabilize and accelerate the training.

\subsection{Results}

In this section, we give a detailed evaluation of the proposed method, including robustness analyses, ablations and comparison to baselines.
We first compare the performance of our models when they are conditioned on different input modality combinations; 
we then perform robustness tests in settings where parts of the modalities are missing as well as when there is a misalignment between video and audio; 
we finally compare to the state-of-the-art on the speaker separation and speech enhancement tasks.

\xpar{Modalities comparison.}
In order to assess the effect of using different modalities as
conditioning input, we compare models using only 
video (\textit{A+V}), only text (\textit{A+T}) or both (\textit{A+V+T}) in \tbl{tab:modalities}.
We observe that the text-only model successfully separates the speakers, obtaining reasonable performance. 
This result is evidence that (a) indeed text can be used to separate speech in cocktail party scenarios, and
(b) that our proposed architecture is flexible enough to capture information from different conditioning sources without any changes and can successfully solve the novel text-conditioned speaker separation task without any help from other modalities.   
Observing the performance of the video-conditioned models shows that video obtains a stronger performance than text.
This is an interesting finding, suggesting that lip movements  are stronger cues for the separation task, presumably because they carry more information than the language content of the utterance (e.g.\ speaker mood, accent etc).  
We refer to \supp for qualitative separation results using the different models.  

\begin{table}[t]
    \setlength{\tabcolsep}{4pt}
    \centering
\begin{tabular}{l|ccc|ccc}
            \toprule
             & \multicolumn{3}{c}{LRS2} & \multicolumn{3}{|c}{LRS3}\\
            Model  & SDR$\uparrow$ & STOI$\uparrow$ & PESQ$\uparrow$ & SDR$\uparrow$ & STOI$\uparrow$ & PESQ$\uparrow$ \\
            \midrule
            A+T & 13.1 & 89.7 & 2.16  &   14.1 & 91.4 & 2.37  \\
            A+V & 14.1 & 91.3 & 2.36  &   \textbf{15.5} & 93.4 & 2.62   \\
            A+V+T  & \textbf{14.2} & \textbf{91.7} & \textbf{2.41}  & \textbf{15.5} & \textbf{93.5} & \textbf{2.63}  \\
            \bottomrule
        \end{tabular}
    \caption{\textbf{Speaker separation performance using different modalities.} 
    We compare \ours models conditioning on different modality combinations. We observe that the text-only model (\textit{A+T}) performs reasonably well for this task, although the performance is lower compared to when conditioning on lip movements from the visual stream.
    The full \textit{A+V+T} model that conditions on both text and video obtains only a slight improvement over the \textit{A+V} model. $\uparrow$ denotes higher is better
    }
    \label{tab:modalities}
    \vspace{-15pt}
\end{table}

\begin{figure}[!htp]
  \subfloat{\includegraphics[width=0.49\linewidth]{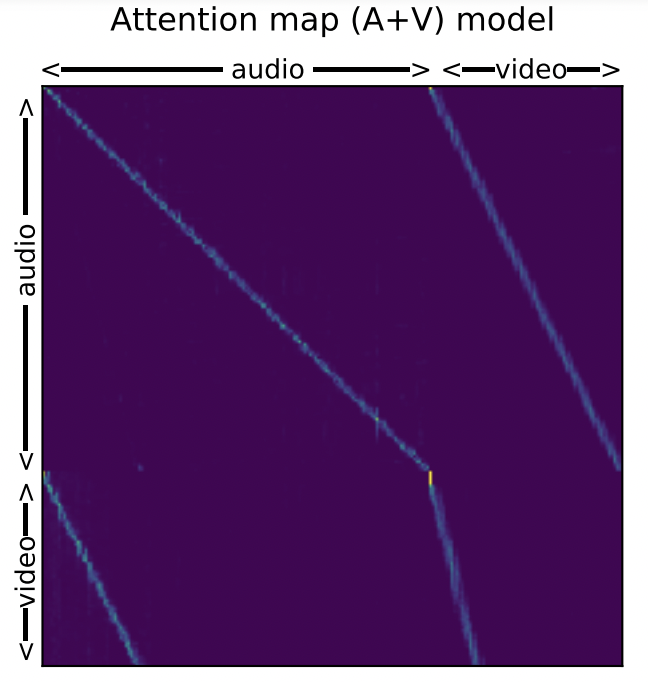}}\hfill
  \subfloat{\includegraphics[width=0.49\linewidth]{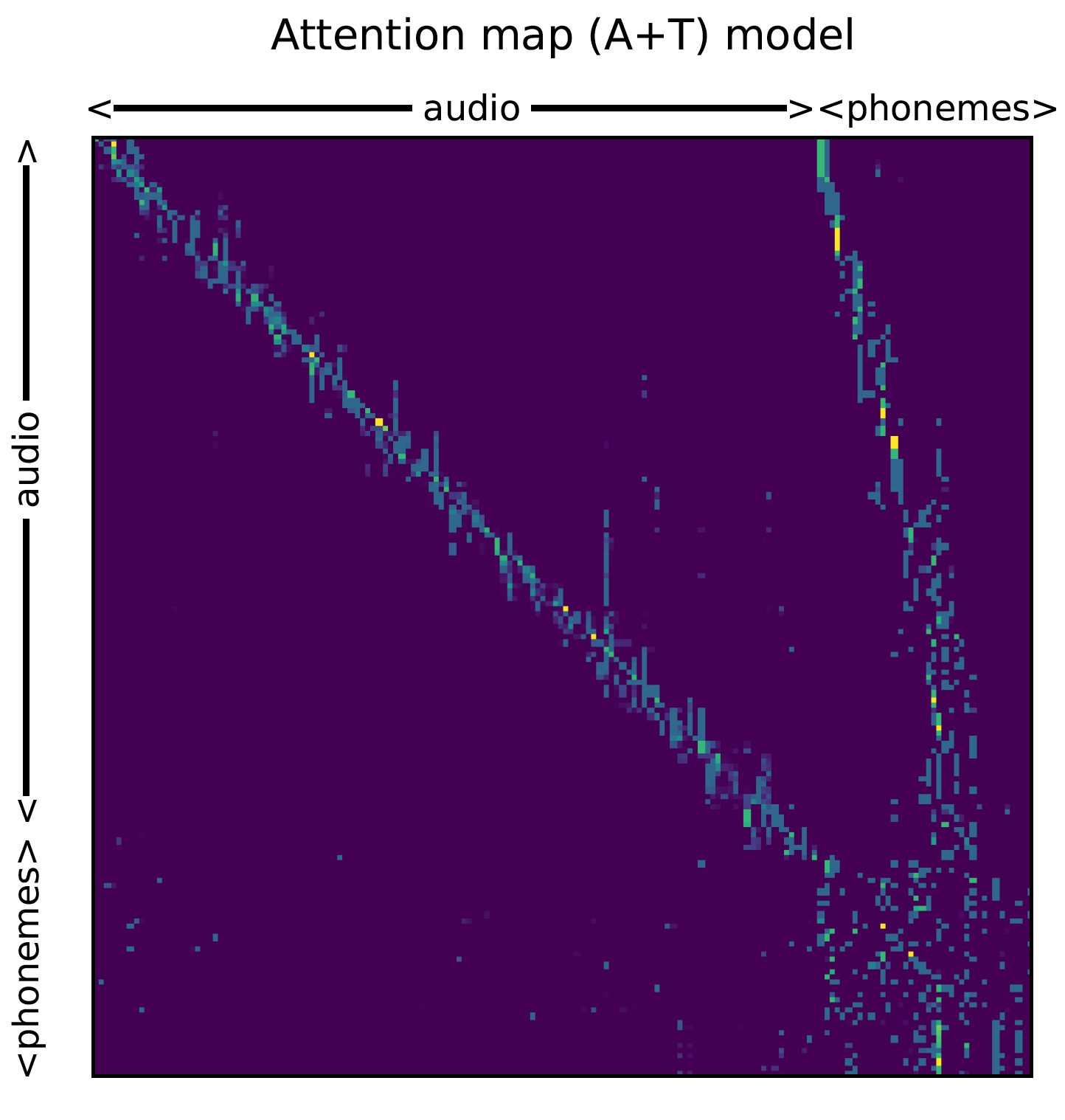}}%
   \caption{\textbf{Attention map visualisations of the first Transformer layer}. The visualisations show the average score of the attention heads in the first multi-head attention layer of the transformer. Brighter colours indicate higher scores and brighter pixels on the same row indicate correspondence between modalities.
   Left: audio and video correlation; and Right:  audio and text correlation. 
  Higher scores are given to the audio token and its corresponding  token in the other modality at each timestep. This indicates that the model is able to elegantly fuses the mixed/noisy audio stream with the conditioning vectors from different modalities, without the need for explicit alignment between the signals, or requiring them to be operating at the same temporal rate. 
}%
   \label{fig:attention_maps}
   \vspace{-5pt}
\end{figure}

\xpar{Cross-modal attention.} To assess the effectiveness of our design for cross-modal attention through a concatenation of the modalities, we examine the attention in the transformer bottleneck. The attention maps in \fig{fig:attention_maps} reveal the correspondence between the audio tokens and the other modalities.  attend to the features in the corresponding modalities. This indicates that the model is able to implicitly learn the audio-visual and audio-text alignments and confirms our intuition that it  does not require manual alignment or common temporal rates.  

\xpar{Architecture component contribution.} To analyse the contribution of our architecture components on the performance of the model, we examine various architecture configurations. The results are shown in~\tbl{tab:lstm-waveform}

\begin{figure}
  \begin{subfigure}{0.9\linewidth}
   \includegraphics[width=\linewidth]{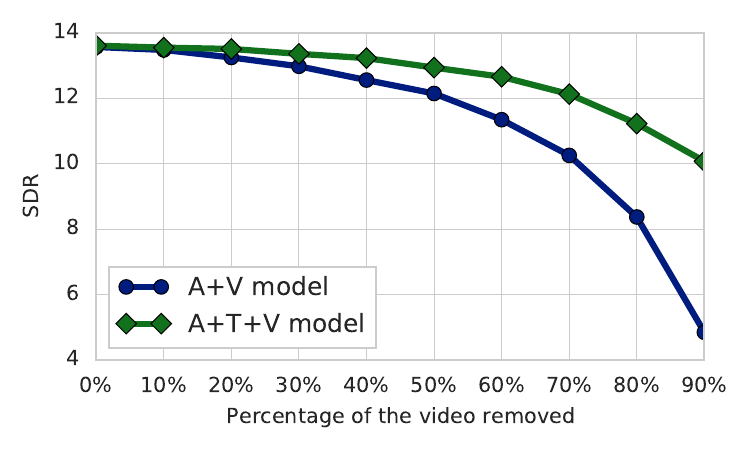}
    \caption{\textbf{Robustness to missing video information.}
    Conditioning on text in combination with video clearly improves the robustness to missing video information (e.g. due to occlusions), compared to only using video.
    }
    \label{fig:video_removed}
  \end{subfigure}
  \hfill
  \begin{subfigure}{0.9\linewidth}
   \includegraphics[width=\linewidth]{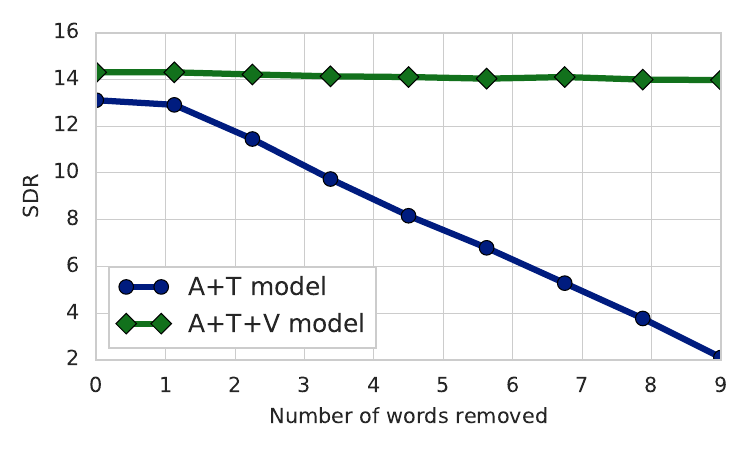}
    \caption{\textbf{Robustness to missing text information.}
    The performance of the text-only conditioned model quickly deteriorates when words are removed from the text input.
    This demonstrates that the separation is indeed reliant on the text content throughout the utterance, and not just for disambiguation between two separated signals.  
    As expected, combining video and text provides robustness to missing text input. 
    }
    \label{fig:text_removed}
  \end{subfigure}
  \vspace{-10pt}
  \caption{Experiments with missing information. 
  }
  \label{fig:short}
  \vspace{-10pt}
\end{figure}

\begin{table}[ht]
    \setlength{\tabcolsep}{4pt}
    \centering
        \begin{tabular}{lcccc}
            \toprule
            Model  & input & SDR$\uparrow$ & STOI$\uparrow$ & PESQ$\uparrow$  \\
            \midrule
            LSTM (A+V)\cite{Afouras20b} & S  & 9.25 & 84.0 & 1.91  \\ 
            UNet + LSTM (A+V)  & W & 12.8 & 89.9 & 2.17  \\ 
             UNet + Transformer (A+V)  & W & 14.1 & 91.3 & 2.36  \\ 
            \bottomrule
        \end{tabular}
    \caption{\textbf{Performance based on various architecture configurations.} We observe that the UNet module, which digests audio as a raw waveform, improves the SDR by 3.55dB in comparison to the spectrograms based baseline model \cite{Afouras20b}. Replacing the LSTM with a transformer bottleneck further increases the SDR by 1.3dB. Note that the UNet removes the necessity to calculate spectrograms and predict phase, while the transformer offers robustness to audio visual misalignment. $\uparrow$ denotes higher is better. S indicates the audio input is in mel-spectrogram form and W indicates raw waveform is used.}
    \label{tab:lstm-waveform}
    \vspace{-10pt}
\end{table}
\xpar{Robustness to missing information.}
As can be seen from the experiments presented so far, although text can be sufficient for performing speech separation, it provides a very limited performance boost when strong visual
evidence (\ie~clear, unoccluded lip movements) is already available. However, we emphasise that adding text in this setting improves the robustness of the model against missing information.
We therefore conduct further experiments where we evaluate the same models, but artificially limit the amount of information from one of the sources. 
For simulating missing video information we mask out (by zeroing) the visual features for a percentage of video frames.
We show the results of these experiments in \fig{fig:video_removed}. 
It is clear that the \textit{A+T+V} model that conditions on text in addition to video is much more robust to distortions in the video input. 

Similarly, for simulating missing text information, we remove a variable number of words from the text input. 
The performance against the number of removed words is shown in \fig{fig:text_removed}. 
We make two observations. First, the performance of the
\textit{A+T} model sharply deteriorates when an increasing number of words are removed from the text input. This indicates that the model relies on the text input to perform the separation of the whole utterance, rather than for just disambiguation between two separated streams.   
Second, the \textit{A+V+T} model is, as expected, very robust to the missing text information. 

\xpar{Robustness to inconsistent modalities.}
To further probe into our models and gain a better understanding of them, we perform a series of
experiments where we replace the conditioning video or text input with input from a different, irrelevant video clip.
The results are shown in \tbl{tab:inconsistent}.
We observe that the models that condition on a single source (\textit{A+T}, \textit{A+V}) completely fail when presented with wrong inputs.
This is expected as the visual/textual evidence in those cases is not consistent with any speech components
contained in the audio.
From a similar analysis of the \textit{A+V+T} model we
observe that, although the model performs a lot worse when supplied with inconsistent video, but consistent text, input, it does not completely fail (\eg obtains $5.1$ SDR). On the other hand, the performance drops only marginally when the model is presented with inconsistent  text but consistent video input.   

To sum up our analysis of the behaviour of \ours, we conclude that 
the \textit{A+V+T} model provides good robustness to disruptions
in the video inputs, which comes with virtually no risk; even if for some reason
the textual input provided is missing or inconsistent with
the video (e.g.\ if we use noisy ASR approximations), the model still performs on par with the video-only model.

\begin{table}[ht]
    \setlength{\tabcolsep}{7pt}
    \centering
        \begin{tabular}{l cc ccc}
            \toprule
            Modality & Text & Video  & SDR$\uparrow$ & STOI$\uparrow$ & PESQ$\uparrow$  \\
            \midrule
            A+T    & \cmark & \xmark  & 13.1 & 89.7 & 2.16   \\
            A+T    & \cuttmark & \xmark  & 1.18 & 60.5 & 1.53  \\
            \midrule
            A+V    & \xmark & \cmark  & 14.1 & 91.3 & 2.36  \\
            A+V    & \xmark & \cuttmark  & -1.51 & 51.9 & 1.39  \\
            \midrule
            A+V+T  & \cmark & \cmark  & 14.2 & 91.7 & 2.41  \\
            A+V+T  & \cmark & \cuttmark  & 5.11 & 70.5 & 1.69 \\
            A+V+T  & \cmark & \xmark & 10.9 & 83.6 & 2.12  \\
            A+V+T  & \cuttmark & \cmark  & 14.1 & 91.3 & 2.36  \\
            A+V+T  & \xmark & \cmark  & 14.1 & 91.2 & 2.34  \\
            \bottomrule
        \end{tabular}
    \caption{
    \textbf{Missing or inconsistent modalities.} 
    \cmark~indicates that the correct signal for the corresponding
    modality is in input, \cuttmark~that the input for this modality is
    supplied from a different sentence or video, and \xmark{} that this
    input is not provided.
    The results are reported for the speaker separation task on the LRS2 test set.
    We observe that the models that use a single modality completely fail to solve the task when the conditioning input is wrong.
    On the other hand, the \textit{A+V+T} model that uses both video and text is fully robust to inconsistent text inputs, and can even partially recover from inconsistent video inputs.  
    $\uparrow$ denotes higher is better}
    \label{tab:inconsistent}
    \vspace{-10pt}
\end{table}

\xpar{Bottleneck ablation and robustness to AV misalignment.}
To assess our choice of the Transformer encoder in the U-Net bottleneck we conduct an experiment
replacing the Transformer with a 2 layer LSTM
(similar to the architecture used by \cite{defossez2020real}),
where the audio and video features are concatenated in the channel dimension.

\begin{figure}
  \centering
   \includegraphics[width=\linewidth]{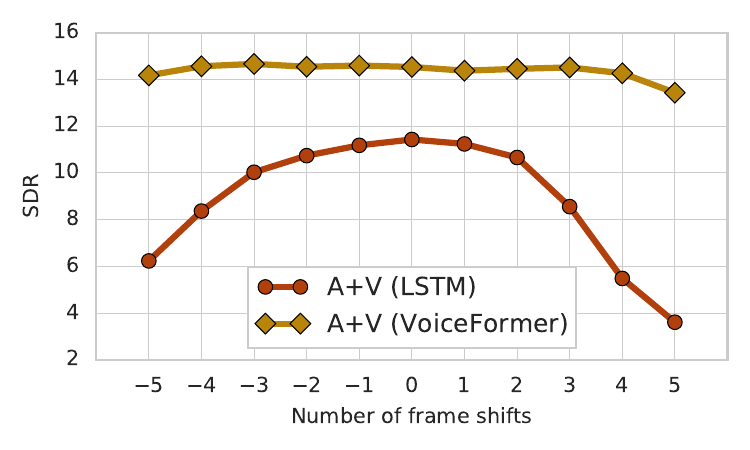}
   \vspace{-25pt}
   \caption{ \textbf{Robustness to audio-visual misalignment}. We compare our proposed model with a baseline using an LSTM bottleneck in the audio-visual speaker separation setting.
   It is clear that while the LSTM baseline struggles when the video
   and audio streams are misaligned, \ours is robust to synchronisation offsets. A  five-frame offset corresponds to 200 ms.
   }
   \label{fig:sync_test}
   \vspace{-10pt}
\end{figure}

We argue that using a Transformer to fuse the different modalities offers the advantage of not requiring synchronised input streams (e.g. video and audio).
To verify this hypothesis, we experiment with artificially shifting the audio input by a random offset within the range from -200 to 200 ms.
We train and evaluate both the LSTM and Transformer models under those conditions. 
The results of this comparison are shown in~\fig{fig:sync_test}.
It is clear that the performance of the LSTM baseline steeply deteriorates
when the audio and video inputs are not properly synchronised. \ours on the other hand is very robust to synchronisation issues, retaining high SDR scores ($>12$), even when the two modalities are offset by up to 200ms.

\begin{table}[ht]
    \small
    \setlength{\tabcolsep}{2pt}
    \centering
        \begin{tabular}{lcc|ccc|ccc}
            \toprule
             &&& \multicolumn{3}{c}{LRS2} & \multicolumn{3}{|c}{LRS3}\\
            Model  & V & T  & SDR & STOI & PESQ & SDR & STOI & PESQ \\
            \midrule
            Noisy input &\xmark &\xmark  & 0.2 & 66.6 & 1.17 &   1.3 & 69.7 & 1.3 \\
            \midrule
            Denoiser~\cite{defossez2020real} &\xmark &\xmark  & 0.2 & 66.6 & 1.17 &   1.3 & 69.7 & 1.3  \\
            AVObjects~\cite{Afouras20b}     &\cmark &\xmark  & 8.86 & 83.9 & 1.94 &   9.72 & 85.1 & 2.02 \\
            Conversation~\cite{Afouras18}  &\cmark &\xmark  & 9.25 & 84.0 & 1.91 &   10.15 & 86.5 & 2.08 \\
            VisualVoice~\cite{gao2021visualvoice}$^\dagger$     &\cmark &\xmark     & 10.8 & 88.4 & 2.16 &   11.7 & 90.0 & 2.41 \\
            Ours  A+V   &\cmark &\xmark     & \underline{14.1} & \underline{91.3} & \underline{2.36} &   \underline{\textbf{15.5}} & \underline{93.4} & \underline{2.62} \\
            \midrule
            Lee \etal\cite{lee2021looking}$^\circ$ &\cmark &\xmark  & 10.01 & 88.0 & 0.94  &   9.78 & 85.0 & 0.710   \\
            Ours  A+V $^\circ$ &\cmark &\xmark  & 12.71 & 92.0 & 2.06  &   14.74 & 94.0 & 2.42   \\
            
            \midrule
            Ours  A+V+T   &\cmark &\cmark   & \textbf{14.2} & \textbf{91.7} & \textbf{2.41} &   \textbf{15.5} & \textbf{93.5} & \textbf{2.63} \\
            \bottomrule
        \end{tabular}
    \caption{\textbf{Comparison to the state-of-the-art on the speaker separation task.}
    We evaluate our best models on the synthetic LRS2 and LRS3 test sets. 
    The \textit{V} and \textit{T} columns denote which modalities are used for conditioning by each model. 
    \textit{A+V+T} indicates our full model and \textit{A+V} for a version conditioning only on video and not on text.
    Our \textit{A+V} model clearly outperforms  the previous work on all the measures, obtaining state-of-the-art performance in this setting.
    We note that the state-of-the-art speech
    enhancement model of~\cite{defossez2020real} cannot deal with a mixture of different speakers and outputs the mixed signal, even after we attempted to fine-tune it on this task.
    $^\dagger$not fine-tuned on the synthetic two-speaker LRS2 training set. $^\circ$ The comparison is on a different test set published by Lee \etal\cite{lee2021looking}. Higher is better for all metrics.
    }
    \label{tab:baselines}
    \vspace{-15pt}
\end{table}

\xpar{Comparison to the state-of-the-art.}
We report our method's performance on speaker separation and compare it to previous methods in \tbl{tab:baselines}.
As baselines, we use the state-of-the-art speech enhancement method of \cite{defossez2020real} (audio-only), as well as the recently proposed audio-visual methods~\cite{Afouras20b,Afouras18,gao2021visualvoice}.
It is clear that \ours outperforms
the previous works on all metrics, obtaining state-of-the-art performance in the (comparable) audio-visual setting \textit{A+V}.

\xpar{Comparison on speech enhancement (denoising).}
For completeness, we also assess our model's performance on the denoising task.
We show the results in \tbl{tab:denoising}.
Our proposed models perform on par with the state-of-the-art Denoiser model~\cite{defossez2020real}.
We note that this is expected as the speech enhancement task is easier than speaker separation and can be solved  by using
only the audio modality.
Indeed the high numbers obtained on all metrics indicate that the performance in this setting is potentially saturated. 
This experiment was included to demonstrate that denoising can also be well handled by the proposed method.
We leave stress-testing of our models on more challenging enhancement settings to future work.  
    
\begin{table}[ht]
    \setlength{\tabcolsep}{8pt}
    \centering
        \begin{tabular}{lccc}
            \toprule
            Model  & SDR$\uparrow$ & STOI$\uparrow$ & PESQ$\uparrow$  \\
            \midrule
            Denoiser~\cite{defossez2020real}  & \textbf{16.49} & 88.4 & 2.44    \\
            \midrule
            Ours A+V     & \underline{16.13} & \textbf{88.9} & \textbf{2.49}  \\
            Ours A+T     & 15.8 & 88.4 & 2.38 \\
            Ours A+V+T   & 16.0 & 88.8 & 2.42  \\
            \bottomrule
        \end{tabular}
    \caption{\textbf{Model performance and comparisons on the speech enhancement (denoising) task.} Results are reported on the LRS2 test set.  All our models obtain good performance, matching the state-of-the-art audio-only~\cite{defossez2019} denoising method. $\uparrow$ denotes higher is better 
    }
    \label{tab:denoising}
    \vspace{-10pt}
\end{table}

\xpar{Qualitative examples.} We strongly encourage the reader to refer to \supp for multiple examples of speaker separation and denoising on real videos, which cover many of the above scenarios. We demonstrate that our model can perform speech separation in challenging real-world scenarios. 

\section{Discussion}

\subsection{Limitations}
The main limitation of the proposed method is the strong assumption that the textual content of the target spoken utterance is available as input to the model during inference. As we have argued in the introduction there are many practical uses where this is a valid assumption (\eg~a prepared conference speech or song lyrics). In our qualitative examples, we also show that it is possible to use related technologies such as ASR to obtain approximate transcriptions which can be used in our method for target speaker separation.

\subsection{Societal impact}
The development of strong multi-modal speech enhancement models opens up exciting opportunities for useful applications, as noted in the introduction. The new method can be used without requiring synchronisation while achieving improved performance. It also provides the opportunity for novel applications, for example, it can use subtitles in films to suppress all background music. However, there are possible malign uses of providing a new method to isolate a speaker from others in terms of surveillance.

In the novel setting that we consider in this paper, however, the natural language content is presumed to be already known (or can be obtained beforehand through other means). 
We therefore argue that this concern does not apply in our setting and that overall the potential benefits from benevolent uses (e.g.\ medical applications and smart hearing aids) outweigh the limited risks.

\subsection{Conclusion}
We have presented a multi-modal speech enhancement method that can condition on multiple non-aligned modalities.
We also introduced text-conditioned speech enhancement as a new
task and showed how our proposed architecture can efficiently solve it.
Our trained models demonstrated state-of-the-art performance in a variety of settings as well as robustness to synchronisation issues and missing information.
In future work, we will consider extending our framework by adding new types of embeddings to condition on modalities such as (i) a particular person (expanding on previous work in \cite{gao2021visualvoice})(ii) the language spoken (e.g.\ to pick a French speaker out of English ones). 

\vspace{10pt}\noindent\textbf{Acknowledgements.}
This work is funded by the UK EPSRC AIMS CDT, the EPSRC Programme Grant VisualAI EP/T028572/1,
and a Google-DeepMind Graduate Scholarship.

\vspace{20pt}
{\small
\bibliographystyle{ieee_fullname}
\bibliography{longstrings,shortstrings,vgg_local,vgg_other,refs_akam,refs}

\begin{thebibliography}{10}\itemsep=-1pt

\bibitem{Afouras18}
Triantafyllos Afouras, Joon~Son Chung, and Andrew Zisserman.
\newblock The conversation: Deep audio-visual speech enhancement.
\newblock In {\em INTERSPEECH}, 2018.

\bibitem{Afouras18d}
Triantafyllos Afouras, Joon~Son Chung, and Andrew Zisserman.
\newblock {LRS3-TED:} a large-scale dataset for visual speech recognition.
\newblock In {\em arXiv preprint arXiv:1809.00496}, 2018.

\bibitem{Afouras19b}
Triantafyllos Afouras, Joon~Son Chung, and Andrew Zisserman.
\newblock My lips are concealed: Audio-visual speech enhancement through obstructions.
\newblock In {\em INTERSPEECH}, 2019.

\bibitem{Afouras20b}
Triantafyllos Afouras, Andrew Owens, Joon~Son Chung, and Andrew Zisserman.
\newblock Self-supervised learning of audio-visual objects from video.
\newblock In {\em Proc. ECCV}, 2020.

\bibitem{Assael16}
Yannis~M. Assael, Brendan Shillingford, Shimon Whiteson, and Nando de Freitas.
\newblock Lipnet: Sentence-level lipreading.
\newblock {\em arXiv:1611.01599}, 2016.

\bibitem{phonimizer}
Mathieu Bernard and Hadrien Titeux.
\newblock Phonemizer: Text to phones transcription for multiple languages in python.
\newblock {\em Journal of Open Source Software}, 6:3958, 12 2021.

\bibitem{Chen21b}
Honglie Chen, Weidi Xie, Triantafyllos Afouras, Arsha Nagrani, Andrea Vedaldi, and Andrew Zisserman.
\newblock Audio-visual synchronisation in the wild.
\newblock In {\em Proc. BMVC}, 2021.

\bibitem{Chen}
Zhuo Chen, Yi Luo, and Nima Mesgarani.
\newblock Deep attractor network for single-microphone speaker separation.
\newblock In {\em 2017 IEEE International Conference on Acoustics, Speech and Signal Processing (ICASSP)}, pages 246--250, 2017.

\bibitem{chung17}
Joon~Son Chung, Andrew Senior, Oriol Vinyals, and Andrew Zisserman.
\newblock Lip reading sentences in the wild.
\newblock In {\em Proc. CVPR}, 2017.

\bibitem{Chung16}
Joon~Son Chung and Andrew Zisserman.
\newblock Lip reading in the wild.
\newblock In {\em Proc. ACCV}, 2016.

\bibitem{chung2020facefilter}
Soo-Whan Chung, Soyeon Choe, Joon~Son Chung, and Hong-Goo Kang.
\newblock Facefilter: Audio-visual speech separation using still images.
\newblock {\em arXiv preprint arXiv:2005.07074}, 2020.

\bibitem{defossez2020real}
Alexandre Defossez, Gabriel Synnaeve, and Yossi Adi.
\newblock Real time speech enhancement in the waveform domain.
\newblock In {\em Interspeech}, 2020.

\bibitem{defossez2019}
Alexandre D{\'{e}}fossez, Nicolas Usunier, L{\'{e}}on Bottou, and Francis~R. Bach.
\newblock {Music Source Separation in the Waveform Domain}.
\newblock {\em CoRR}, abs/1911.13254, 2019.

\bibitem{ephrat2018looking}
Ariel Ephrat, Inbar Mosseri, Oran Lang, Tali Dekel, Kevin Wilson, Avinatan Hassidim, William Freeman, and Michael Rubinstein.
\newblock Looking to listen at the cocktail party: A speaker-independent audio-visual model for speech separation.
\newblock {\em arXiv preprint arXiv:1804.03619}, 2018.

\bibitem{fisher2000learning}
John~W Fisher~III, Trevor Darrell, William~T Freeman, and Paul~A Viola.
\newblock Learning joint statistical models for audio-visual fusion and segregation.
\newblock In {\em NeurIPS}, 2000.

\bibitem{gabbay2018seeing}
Aviv Gabbay, Ariel Ephrat, Tavi Halperin, and Shmuel Peleg.
\newblock Seeing through noise: Visually driven speaker separation and enhancement.
\newblock In {\em Proc. ICASSP}, pages 3051--3055. IEEE, 2018.

\bibitem{gabbay17b}
Aviv Gabbay, Asaph Shamir, and Shmuel Peleg.
\newblock {Visual Speech Enhancement using Noise-Invariant Training}.
\newblock {\em arXiv preprint arXiv:1711.08789}, 2017.

\bibitem{Gan2020MusicGF}
Chuang Gan, Deng Huang, Hang Zhao, Joshua~B. Tenenbaum, and Antonio Torralba.
\newblock Music gesture for visual sound separation.
\newblock {\em 2020 IEEE/CVF Conference on Computer Vision and Pattern Recognition (CVPR)}, pages 10475--10484, 2020.

\bibitem{gao18graum}
Ruohan Gao, Rog{\'{e}}rio~Schmidt Feris, and Kristen Grauman.
\newblock Learning to separate object sounds by watching unlabeled video.
\newblock {\em CoRR}, abs/1804.01665, 2018.

\bibitem{gao2019co}
Ruohan Gao and Kristen Grauman.
\newblock Co-separating sounds of visual objects.
\newblock {\em arXiv preprint arXiv:1904.07750}, 2019.

\bibitem{gao2021visualvoice}
Ruohan Gao and Kristen Grauman.
\newblock {VisualVoice: Audio-Visual Speech Separation with Cross-Modal Consistency}.
\newblock In {\em Proc. CVPR}, 2021.

\bibitem{gu2020multi}
Rongzhi Gu, Shi-Xiong Zhang, Yong Xu, Lianwu Chen, Yuexian Zou, and Dong Yu.
\newblock Multi-modal multi-channel target speech separation.
\newblock {\em IEEE Journal of Selected Topics in Signal Processing}, 14(3):530--541, 2020.

\bibitem{Hershey}
John~R. Hershey, Zhuo Chen, Jonathan Le~Roux, and Shinji Watanabe.
\newblock Deep clustering: Discriminative embeddings for segmentation and separation.
\newblock In {\em 2016 IEEE International Conference on Acoustics, Speech and Signal Processing (ICASSP)}, pages 31--35, 2016.

\bibitem{jaegle2021perceiver}
Andrew Jaegle, Felix Gimeno, Andrew Brock, Andrew Zisserman, Oriol Vinyals, and Joao Carreira.
\newblock Perceiver: General perception with iterative attention.
\newblock {\em arXiv preprint arXiv:2103.03206}, 2021.

\bibitem{lee2021looking}
Jiyoung Lee, Soo-Whan Chung, Sunok Kim, Hong-Goo Kang, and Kwanghoon Sohn.
\newblock Looking into your speech: Learning cross-modal affinity for audio-visual speech separation.
\newblock In {\em Proceedings of the IEEE/CVF Conference on Computer Vision and Pattern Recognition (CVPR)}, 2021.

\bibitem{lee2021parameter}
Sangho Lee, Youngjae Yu, Gunhee Kim, Thomas Breuel, Jan Kautz, and Yale Song.
\newblock Parameter efficient multimodal transformers for video representation learning.
\newblock In {\em Proc. ICLR}, 2021.

\bibitem{Lin_2020_ACCV}
Yan-Bo Lin and Yu-Chiang~Frank Wang.
\newblock Audiovisual transformer with instance attention for audio-visual event localization.
\newblock In {\em Proc. ACCV}, 2020.

\bibitem{Luo2018}
Yi Luo, Zhuo Chen, and Nima Mesgarani.
\newblock Speaker-independent speech separation with deep attractor network.
\newblock {\em IEEE/ACM Transactions on Audio, Speech, and Language Processing}, 26(4):787--796, 2018.

\bibitem{Moray59cocktailparty}
Neville Moray.
\newblock Attention in dichotic listening: Affective cues and the influence of instructions.
\newblock {\em Quarterly Journal of Experimental Psychology}, 11(1):56--60, 1959.

\bibitem{nagrani2021attention}
Arsha Nagrani, Shan Yang, Anurag Arnab, Aren Jansen, Cordelia Schmid, and Chen Sun.
\newblock Attention bottlenecks for multimodal fusion.
\newblock {\em NeurIPS}, 2021.

\bibitem{owens2018b}
Andrew Owens and Alexei~A. Efros.
\newblock Audio-visual scene analysis with self-supervised multisensory features.
\newblock {\em Proc. ECCV}, 2018.

\bibitem{Parekh17}
Sanjeel Parekh, Slim Essid, Alexey Ozerov, Ngoc Q.~K. Duong, Patrick Pérez, and Gaël Richard.
\newblock Motion informed audio source separation.
\newblock In {\em 2017 IEEE International Conference on Acoustics, Speech and Signal Processing (ICASSP)}, pages 6--10, 2017.

\bibitem{Jie17}
Jie Pu, Yannis Panagakis, Stavros Petridis, and Maja Pantic.
\newblock Audio-visual object localization and separation using low-rank and sparsity.
\newblock In {\em 2017 IEEE International Conference on Acoustics, Speech and Signal Processing (ICASSP)}, pages 2901--2905, 2017.

\bibitem{Raffel14mireval}
Colin Raffel, Brian Mcfee, Eric Humphrey, Justin Salamon, Oriol Nieto, Dawen Liang, and Daniel Ellis.
\newblock mir\_eval: A transparent implementation of common mir metrics.
\newblock In {\em Proceedings - 15th International Society for Music Information Retrieval Conference (ISMIR 2014)}, 10 2014.

\bibitem{reddy2021interspeech}
Chandan~KA Reddy, Harishchandra Dubey, Kazuhito Koishida, Arun Nair, Vishak Gopal, Ross Cutler, Sebastian Braun, Hannes Gamper, Robert Aichner, and Sriram Srinivasan.
\newblock Interspeech 2021 deep noise suppression challenge.
\newblock In {\em INTERSPEECH}, 2021.

\bibitem{Prajwal21}
Prajwal~K Renukanand, Liliane Momeni, Triantafyllos Afouras, and Andrew Zisserman.
\newblock Visual keyword spotting with attention.
\newblock In {\em Proc. BMVC}, 2021.

\bibitem{rix2001perceptual}
Antony~W Rix, John~G Beerends, Michael~P Hollier, and Andries~P Hekstra.
\newblock Perceptual evaluation of speech quality (pesq)-a new method for speech quality assessment of telephone networks and codecs.
\newblock In {\em Proc. ICASSP}, volume~2, pages 749--752. IEEE, 2001.

\bibitem{ronneberger2015u}
Olaf Ronneberger, Philipp Fischer, and Thomas Brox.
\newblock U-net: Convolutional networks for biomedical image segmentation.
\newblock In {\em International Conference on Medical image computing and computer-assisted intervention}, pages 234--241. Springer, 2015.

\bibitem{rouditchenko2019self}
Andrew Rouditchenko, Hang Zhao, Chuang Gan, Josh McDermott, and Antonio Torralba.
\newblock Self-supervised audio-visual co-segmentation.
\newblock In {\em Proc. ICASSP}, pages 2357--2361. IEEE, 2019.

\bibitem{sadeghi2020audio}
Mostafa Sadeghi, Simon Leglaive, Xavier Alameda-Pineda, Laurent Girin, and Radu Horaud.
\newblock Audio-visual speech enhancement using conditional variational auto-encoders.
\newblock {\em IEEE/ACM Transactions on Audio, Speech, and Language Processing}, 28:1788--1800, 2020.

\bibitem{schwartz2004b69}
Jean-Luc Schwartz, Fr{\'e}d{\'e}ric Berthommier, and Christophe Savariaux.
\newblock Seeing to hear better: evidence for early audio-visual interactions in speech identification.
\newblock {\em Cognition}, 93(2):B69--B78, 2004.

\bibitem{Stafylakis17}
Themos Stafylakis and Georgios Tzimiropoulos.
\newblock Combining residual networks with lstms for lipreading.
\newblock In {\em Interspeech}, 2017.

\bibitem{taal11}
Cees Taal, Richard Hendriks, Richard Heusdens, and Jesper Jensen.
\newblock An algorithm for intelligibility prediction of time-frequency weighted noisy speech.
\newblock {\em IEEE Transactions on Audio, Speech and Language Processing}, 2011.

\bibitem{tian2020avvp}
Yapeng Tian, Dingzeyu Li, and Chenliang Xu.
\newblock Unified multisensory perception: Weakly-supervised audio-visual video parsing.
\newblock In {\em Proc. ECCV}, 2020.

\bibitem{tzinis2020into}
Efthymios Tzinis, Scott Wisdom, Aren Jansen, Shawn Hershey, Tal Remez, Daniel~PW Ellis, and John~R Hershey.
\newblock Into the wild with audioscope: Unsupervised audio-visual separation of on-screen sounds.
\newblock {\em arXiv preprint arXiv:2011.01143}, 2020.

\bibitem{Tzinis2021ImprovingOS}
Efthymios Tzinis, Scott Wisdom, Tal Remez, and John~R. Hershey.
\newblock Improving on-screen sound separation for open domain videos with audio-visual self-attention.
\newblock {\em ArXiv}, abs/2106.09669, 2021.

\bibitem{fevotte05}
Emmanuel Vincent, Rémi Gribonval, and Cédric Févotte.
\newblock {BSS EVAL} toolbox user guide.
\newblock {\em IRISA Technical Report 1706. http://www.irisa.fr/metiss/bss eval/.}, 2005.

\bibitem{Wang}
DeLiang Wang and Jitong Chen.
\newblock Supervised speech separation based on deep learning: An overview.
\newblock {\em IEEE/ACM Transactions on Audio, Speech, and Language Processing}, 26(10):1702--1726, 2018.

\bibitem{wang2018voicefilter}
Quan Wang, Hannah Muckenhirn, Kevin Wilson, Prashant Sridhar, Zelin Wu, John Hershey, Rif~A Saurous, Ron~J Weiss, Ye Jia, and Ignacio~Lopez Moreno.
\newblock Voicefilter: Targeted voice separation by speaker-conditioned spectrogram masking.
\newblock In {\em Interspeech}, 2018.

\bibitem{Wang2019}
Zhong-Qiu Wang and DeLiang Wang.
\newblock Combining spectral and spatial features for deep learning based blind speaker separation.
\newblock {\em IEEE/ACM Transactions on Audio, Speech, and Language Processing}, 27(2):457--468, 2019.

\bibitem{Wu}
Jian Wu, Yong Xu, Shi-Xiong Zhang, Lian-Wu Chen, Meng Yu, Lei Xie, and Dong Yu.
\newblock Time domain audio visual speech separation.
\newblock In {\em 2019 IEEE Automatic Speech Recognition and Understanding Workshop (ASRU)}, pages 667--673, 2019.

\bibitem{xu2019multilevel}
Huijuan Xu, Kun He, Bryan~A Plummer, Leonid Sigal, Stan Sclaroff, and Kate Saenko.
\newblock Multilevel language and vision integration for text-to-clip retrieval.
\newblock In {\em AAAI}, 2019.

\bibitem{Xu19plusnet}
Xudong Xu, Bo Dai, and Dahua Lin.
\newblock Recursive visual sound separation using minus-plus net.
\newblock In {\em 2019 IEEE/CVF International Conference on Computer Vision (ICCV)}, pages 882--891, 2019.

\bibitem{Yu}
Dong Yu, Morten Kolbæk, Zheng-Hua Tan, and Jensen Jensen.
\newblock Permutation invariant training of deep models for speaker-independent multi-talker speech separation.
\newblock In {\em 2017 IEEE International Conference on Acoustics, Speech and Signal Processing (ICASSP)}, pages 241--245, 2017.

\bibitem{yuan2019find}
Yitian Yuan, Tao Mei, and Wenwu Zhu.
\newblock To find where you talk: Temporal sentence localization in video with attention based location regression.
\newblock In {\em AAAI}, 2019.

\bibitem{zhao2019sound}
Hang Zhao, Chuang Gan, Wei-Chiu Ma, and Antonio Torralba.
\newblock The sound of motions.
\newblock {\em Proc. ICCV}, 2019.

\bibitem{zhao2018sound}
Hang Zhao, Chuang Gan, Andrew Rouditchenko, Carl Vondrick, Josh McDermott, and Antonio Torralba.
\newblock The sound of pixels.
\newblock {\em arXiv preprint arXiv:1804.03160}, 2018.

\end{thebibliography}
}
\end{document}


\title{Supplementary Material \\ Reading to Listen at the Cocktail Party: \\ Multi-Modal Speech Separation}

\maketitle

\section{Qualitative examples}
Please visit \small{\url{https://www.robots.ox.ac.uk/~vgg/research/voiceformer}} to see qualitative results of our model on real-world speaker separation and speech enhancement examples.

\section{Architecture details}

As outlined in Section 3.1 of the main paper, the architecture of the
our full model consists of four main parts:
(i) the U-Net audio encoder and decoder,
(ii) the video backbone,
(iii) the text embedding, 
and (iv) the bottlenck Transformer.

\xpar{Visual backbone.}
We provide details of the visual CNN backbone architecture in~\tbl{tab:cnn_arch}.

\xpar{Text embedding.}
The text input is pre-processed to phoneme indexes using phonimizer library \cite{phonimizer}. The phonimizer allows simple phonimization of words using a number of backends. in our work \textit{espeak-eng} was used as the backend to generated the phonemes based on Intenational Phonetic Alphabet. The phonemes are then passed through a learnable embedding of dimension $768$, to match the visual and audio embeding channels. 

\xpar{Audio U-Net encoder-decoder}
For the U-Net model, we directly adopt the architecture of the Denoiser~\cite{defossez2020real}, which is a multi-layer convolutional encoder and decoder with skip connections. The full architecture is shown in \tbl{tab:unet}.

\xpar{Transformer bottleneck.}
The audio, video and text representations are concatenated in time dimension and fed into the transformer network which consists of 3 encoder layers, each layer containing 8 heads with a model size of 532 and no masking has been applied. 

The  output  of  the  Transformer  corresponding to its audio input is passed to the U-net decoder network.

\begin{table}[h]
\setlength{\tabcolsep}{2.0pt}
  \footnotesize
  \centering
  \begin{tabular}[t]{  l r c c c r }
  \toprule
  Layer & \# Channels & Kernel & Stride & Padding  & Output dims  \\  
  \midrule
  video input  & 3 &   - &  -      &   -    & $T_v \times 112 \times 112$  \\  
  conv\textsubscript{1,1}  & 64 & (5,5,5) & (1,2,2) & (2,2,2)  & $T_v \times 56 \times 56 $   \\   
  \midrule
  conv\textsubscript{2,1} & 128 & (3,3)   & (2,2) &  (1,1)  &   $T_v \times 28 \times 28 $  \\ 
  conv\textsubscript{2,2} & 128 & (3,3)   & (1,1) &  (1,1)  &   $T_v \times 28 \times 28 $  \\ 
   conv\textsubscript{2,3} & 128 & (3,3)   & (1,1) &  (1,1)  &   $T_v \times 28 \times 28 $  \\ 
  \midrule
  conv\textsubscript{3,1} & 256 & (3,3)   & (2,2) &  (1,1)  &   $T_v \times 14 \times 14 $   \\ 
  conv\textsubscript{3,2} & 256 & (3,3)   & (1,1) &  (1,1)  &   $T_v \times 14 \times 14 $  \\ 
  conv\textsubscript{3,3} & 256 & (3,3)   & (1,1) &  (1,1)  &   $T_v \times 14 \times 14 $  \\ 
  \midrule
  conv\textsubscript{4,1} & 512 & (3,3)   & (2,2) &  (1,1)  &   $T_v \times 7 \times 7 $  \\ 
  conv\textsubscript{4,2} & 512 & (3,3)   & (1,1) &  (1,1)  &   $T_v \times 7 \times 7 $  \\ 
  conv\textsubscript{4,3} & 512 & (3,3)   & (1,1) &  (1,1)  &   $T_v \times 7 \times 7 $  \\ 
  \midrule
  conv\textsubscript{5,1} & 512 & (3,3)   & (2,2) &  (1,1)  &   $T_v \times 4 \times 4 $  \\ 
  fc$_1$             & 512 & (4,4)  & (1,1) &  (0,0) & $T_v \times 1 \times 1 $  \\ 
  \bottomrule
  \end{tabular}
  
  \caption{
    Architecture details for the visual CNN backbone.
    Batch Normalization and ReLU activation are added after every convolutional layer. 
    Shortcut connections are also added at each layer, except for the first layer of every residual block -- i.e. the ones with stride $> 1$. 
    The fully connected layers are followed by normalisation layer and a ReLu activation.  
  }
  \label{tab:cnn_arch}
\end{table}

\begin{table}[h]
    \setlength{\tabcolsep}{10pt}
    \centering
        \begin{tabular}{lcccc}
            \toprule
            Model & \# layers  & SDR & STOI & PESQ  \\
            \midrule
            A+V & 2  & 13.6 & 89.7 & 2.23  \\ 
            A+V & 3  & 14.1 & 91.3 & 2.36  \\ 
            \midrule
            A+V+T & 2  & 13.8 & 91.1 & 2.31  \\ 
            A+V+T & 3  & 14.2 & 91.7 & 2.41  \\ 
            \bottomrule
        \end{tabular}

    \caption{\textbf{Architecture ablations}
    These results are from a balanced LRS2 test set.
    }
    \label{tab:arch_ablations}
\end{table}

\begin{table}[h]
    \setlength{\tabcolsep}{10.0pt}
    \centering
        \begin{tabular}{lcccc}
            \toprule
            Model   & SDR & STOI & PESQ  \\
            \midrule
        
            A+T (LSTM) & 11.2  & 86.3 & 1.95  \\ 
            A+T (Transformer)  & 13.1 & 89.7 & 2.16  \\ 
            \bottomrule
        \end{tabular}
    \caption{\textbf{Performance based on various architecture configurations.} The results show the performance improvements as a result of introducing UNet and Transformer components.}
    \label{tab:lstm-waveform}
  
\end{table}

\begin{table}[ht]
\setlength{\tabcolsep}{5.0pt}
  \footnotesize
  \centering
  \begin{tabular}[t]{  l r c c c r }
  \toprule
  Layer & \# Channels & Kernel & Stride & Padding  & Output dims  \\  
  \midrule
  audio input  & 1 &   - &  -      &   -    & $T_a$  \\  
  \midrule
  conv$^e_{1,1}$ & 48 & 8 & 4 & 2  &   $ \nicefrac{T_a}{4}  $  \\ 
  conv$^e_{1,2}$ & 48 & 1 & 1 & 0  &   $ \nicefrac{T_a}{4}  $  \\ 
  \midrule
  conv$^e_{1,1}$ & 96 & 8 & 4 & 2  &   $ \nicefrac{T_a}{16}  $  \\ 
  conv$^e_{1,1}$ & 96 & 1 & 1 & 0  &   $ \nicefrac{T_a}{16}  $  \\ 
  \midrule
  conv$^e_{3,1}$ & 192 & 8 & 4 & 2  &   $ \nicefrac{T_a}{256}  $  \\ 
  conv$^e_{3,2}$ & 192 & 1 & 1 & 0  &   $ \nicefrac{T_a}{256}  $  \\ 
  \midrule
  conv$^e_{4,1}$ & 384 & 8 & 4 & 2  &   $ \nicefrac{T_a}{1024}  $  \\ 
  conv$^e_{4,2}$ & 384 & 1 & 1 & 0  &   $ \nicefrac{T_a}{1024}  $  \\ 
  \midrule
  conv$^e_{5,1}$ & 768 & 8 & 4 & 2  &   $ \nicefrac{T_a}{4096}  $  \\ 
  conv$^e_{5,2}$ & 768 & 1 & 1 & 0  &   $ \nicefrac{T_a}{4096}  $  \\ 
  \midrule
  bottleneck     & 768 & - & - & -  &   $ \nicefrac{T_a}{4096}  $  \\ 
  \midrule
  conv$^e_{5,2}$ & 768 & 1 & 1 & 0  &   $ \nicefrac{T_a}{4096}  $  \\ 
  conv$^e_{5,1}$ & 384 & 8 & \nicefrac{1}{4} & 2  &   $ \nicefrac{T_a}{1024}  $  \\ 
  \midrule
  conv$^e_{4,2}$ & 384 & 1 & 1 & 0  &   $ \nicefrac{T_a}{1024}  $  \\ 
  conv$^e_{4,1}$ & 192 & 8 & \nicefrac{1}{4} & 2  &   $ \nicefrac{T_a}{256}  $  \\ 
  \midrule
  conv$^e_{3,2}$ & 192 & 1 & 1 & 0  &   $ \nicefrac{T_a}{256}  $  \\ 
  conv$^e_{3,1}$ & 48 & 8 & \nicefrac{1}{4} & 2  &   $ \nicefrac{T_a}{4}  $  \\ 
  \midrule
  conv$^e_{1,2}$ & 48 & 1 & 1 & 0  &   $ \nicefrac{T_a}{4}  $  \\ 
  conv$^e_{1,1}$ & 1 & 8 & \nicefrac{1}{4} & 2  &   $ T_a  $  \\ 
  \midrule
  audio output  & 1 &   - &  -      &   -    & $T_a$  \\  
 
  \bottomrule
  \end{tabular}
 
  \caption{
    Architecture details for the audio  U-Net encoder-decoder.
    ReLu activations are added after every strided convolution layer for the encoder and decoder blocks. 
    GLU activations are added after every first $1 \times 1$ convolutional layer of both encoder and decoder blocks. 
    A skip connection connects the output of the i-th layer of the encoder and the input of the i-th layer of the decoder. 
    \textit{bottleneck} refers to the output of the Transformer bottleneck that has the same temporal dimensions as the audio  embeddings.  
    For the last decoder layer the output is a single channel and there is no ReLU activation.
  }
  \label{tab:unet}
\end{table}
\section{Architecture ablations}

In order to assess the affect of the bottleneck model capacity on its performance we also perform an ablations varying the number of Transformer encoder layers used by our models. The results of this analysis are shown in \tbl{tab:arch_ablations}. We notice a small increase in
performance when using 3 instead of 2 Transformer layers. This shows the importance of the bottleneck layer and that increasing its capacity can lead to further performance boosts.


We also trained a model that ingests text with an LSTM, instead of the transformer, using linearly interpolated in time word embeddings. The results are shown in~\tbl{tab:lstm-waveform}.
It is clear that \ours outperforms the LSTM baseline by a large margin (\eg~ -- relative SDR improvement). 
We also emphasise that adding text to the transformer model is more elegant and natural in comparison to using the LSTM for this task.


\vspace{20pt}
{\small
\bibliographystyle{ieee_fullname}
\bibliography{longstrings,shortstrings,vgg_local,vgg_other,refs_akam,refs}
}